\documentclass[preprint,showpacs,preprintnumbers,amsmath,amssymb]{revtex4}
\newtheorem{theorem}{Theorem}
\newtheorem{definition}{Definition}
\newtheorem{lemma}{Lemma}

\newtheorem{example}{Example}

\usepackage{graphicx}
\usepackage{dcolumn}
\usepackage{bm}

\begin{document}
\title{
No-Cloning Theorem on Quantum Logics
}

\author{Takayuki Miyadera$\ ^*$
and Hideki Imai$\ ^{*,\dagger}$
%
\\
$\ ^*$
Research Center for Information Security (RCIS), \\
National Institute of Advanced Industrial \\ 
Science and Technology (AIST). \\
Daibiru building 1003,\\
Sotokanda, Chiyoda-ku, Tokyo, 101-0021, Japan.
\\
(e-mail: miyadera-takayuki@aist.go.jp)
\\
$\ ^{\dagger}$
Graduate School of Science and Engineering,
\\
Chuo University. \\
1-13-27 Kasuga, Bunkyo-ku, Tokyo 112-8551, Japan .
}

\begin{abstract}
This paper discusses the
no-cloning theorem in a logico-algebraic approach. 
In this approach, an orthoalgebra is considered as
a general structure for 
propositions in a physical theory. 
We proved that an orthoalgebra admits cloning operation 
if and only if
it is a Boolean algebra. 
That is, only classical theory 
admits the cloning of states.
If unsharp propositions are to be included in the theory, 
then a notion of effect algebra is considered. 
We proved that an atomic Archimedean 
effect algebra admitting cloning operation is
a Boolean algebra. 
This paper also presents a partial result indicating a relation 
between cloning on effect algebras and hidden variables.
\end{abstract}
\maketitle
\section{Introduction}
\label{intro}
In 1982, Wootters and Zurek \cite{WZ} and Dieks \cite{Dieks} elucidated the
no-cloning theorem: unknown quantum states cannot be cloned.
The no-go theorem that prohibits the universal cloning of quantum states
often plays a central role in quantum 
information \cite{Nielsen}. 
For instance, 
in quantum cryptography \cite{BB84}, 
applying this theorem, legitimate users could detect 
an eavesdropper who pilfered the information. 
Recently, Barnum, Barrett, Leifer and Wilce \cite{Barnum}
reported a generalization of the theorem 
in general probability theory (or the convex approach),  
the framework of which is sufficiently broad to
treat both classical theory and quantum theory as 
its examples. 
 The structure of state space is considered 
 a fundamental object for this framework. 
 For instance, a classical theory 
 can be characterized by its convex state space 
 being a simplex.
 Barnum, Barrett, Leifer and Wilce 
 elucidated that if the state of a system can be cloned, 
 then the state space is a simplex.
This paper discusses an approach to address
 this problem 
in the context of quantum logics. 
This approach was originated by Birkhoff and von Neumann \cite{vN}
in 1936. 
%
Here, 
the structure of propositions is
the most fundamental object. 
A theory is determined by specifying 
 an algebra consisting of the propositions.
 A classical theory is 
identified with a Boolean algebra. 
Birkhoff and von Neumann studied properties 
that are
satisfied by projection operators 
in Hilbert space. 
The proposition system of quantum theory,
in contrast to classical theory, 
does not satisfy the distributive law and thus
is not identified with a Boolean algebra.
Nonetheless, it
satisfies the weaker axioms for an orthomodular lattice. 
Since then, numerous studies have been 
conducted to justify the common Hilbert space formalism 
of quantum mechanics (see \cite{TheBook} and 
references therein). 
A typical study in this
direction
starts with a very general algebra and then 
certain reasonable conditions are imposed on it. 
In this approach, orthoalgebra is considered
 as a general structure; 
the Boolean algebra and the orthomodular 
lattice are examples for the same. 
If unsharp propositions are to be included in the theory, 
then a notion of effect algebra is considered \cite{FoBe1}.
The objective of this paper is to assess the
cloning process on the above-mentioned algebras and the
 conditions required for them to satisfy the 
no-cloning theorem.
\par
This paper is organized as follows: 
Section \ref{sect:ortho} presents 
a brief review of orthoalgebra.
Section \ref{sect:main} provides our main result 
that can be regarded as a no-cloning theorem on the orthoalgebras.
Here we prove that if an orthoalgebra admits cloning it is definitely
a Boolean algebra. Hence, only classical theory admits 
cloning. This result agrees with the result obtained 
by Barnum, Barrett, Leifer and Wilce in general probability theory. 
Section \ref{sect:effect} presents 
partial results of extension of the earlier result to effect algebras. 
\section{Orthoalgebras}\label{sect:ortho}
An orthoalgebra, consisting of 
sharp propositions, is a generalized structure of 
the Boolean algebra 
and the orthomodular lattice,
 which play an important role in 
the investigation of quantum logic. 
Its definition is stated as follows: 
\cite{TheBook}
\begin{definition}\label{def:ortho}
Let us consider $(P,0,1,\oplus)$ consisting of a set 
$P$ which contains two special elements $0$ and $1$ and 
a partially 
defined binary operation $\oplus$.
If 
the quadruple 
satisfies the following conditions for 
all $p, q, r\in P$, 
then $(P,0,1,\oplus)$ is called an orthoalgebra. 
\begin{itemize}
\item[(i)]
If $p\oplus q$ is defined (denoted as $p \perp q$), then 
$q\oplus p$ is also defined and $p\oplus q =q\oplus p$ holds. 
\item[(ii)]
If $q\perp r$ and $p\perp (q\oplus r)$ hold, 
then $p\perp q$ and $(p\oplus q) \perp r$, and
$p\oplus (q\oplus r)=(p\oplus q) \oplus r$ hold. 
\item[(iii)]
 For every $p\in P$, there exists a unique $q\in P$ 
 such that $p\perp q$ and $p\oplus q =1$ hold. 
 We represent such uniquely determined $q$ as $p'$. 
 \item[(iv)]
 If $p\perp p$, then $p=0$.  
\end{itemize}
\end{definition}
\begin{example}
A simple example is the set of projection operators 
in Hilbert space. Let ${\cal H}$ be a Hilbert space and 
$P({\cal H})$ be the set of all the projection operators on it.
In $P({\cal H})$, a partially defined binary operation
is introduced by 
$p\oplus q=p+q$ (summation of operators), for $p,q \in P({\cal H})$ 
with $pq=0$. 
$0$ and $1$ are a null operator and an identity operator on 
${\cal H}$, respectively. 
Hence from the above definition, 
it can be derived that $(P({\cal H}),
0,1,\oplus)$ becomes an orthoalgebra. 
\end{example} 
A partial order in 
$(P,0,1,\oplus)$ can be introduced. 
\begin{definition}
If there exists an element $r\in P$ such that 
$p\perp r$ and $q=p\oplus r$ hold, 
we denote as $p\leq q$ (or equivalently $q\geq p$).
\end{definition}
It can be confirmed that $\leq$ forms 
a partial order and satisfies
$0\leq p \leq 1$ 
for every element $p\in P$.
That is, $(P,0,1,\leq)$ forms a bounded poset. 
We denote 
the least upper and the greatest lower bounds of
$\{p,q\}$
by $p\vee q$ and 
$p\wedge q$ (unique), respectively, 
if they exist. 
If $P$ is an 
orthoalgebra, it can be proved 
that $p\wedge p'=0$ for any element $p\in P$. 
If $p\perp q$ and $p\vee q$ exists, 
it coincides with 
$p\oplus q$. 
An orthomodular poset may be defined as an 
orthoalgebra $P$ such that the coherence law is 
satisfied. That is, for any mutually orthogonal 
$p,q$ and $r\in P$, $(p\oplus q)\oplus r$ is defined. 
Two elements $p,q \in P$ are said to be compatible 
if there exist mutually orthogonal elements $x,y$ and $z$ satisfying 
$p=x\oplus z$ and $q=y\oplus z$. 
If an orthomodular poset $P$ satisfies 
the compatibility condition, 
that is, every pair $p,q\in P$ is 
compatible, then $P$ becomes a Boolean algebra. 
\par
Having defined an orthoalgebra 
$P$ which is a set of propositions, 
we can now introduce states and dynamics on it. 
\begin{definition}
Let $P$ be an orthoalgebra. 
A state on $P$ is a map $\mu: P\to {\bf R}$
such that, 
for any $p,q \in P$ with $p\perp q$, 
$\mu(p\oplus q)=\mu(p)+\mu(q)$ holds, and
$\mu(p)\geq 0$ for any $p$ and $\mu(1)=1$ are satisfied. 
\end{definition}
A nonnegative value $\mu(p)$ is interpreted as the 
probability to obtain `Yes' when a measurement 
of $p$ for the state $\mu$ is made.
We assume a sufficient number of states on orthoalgebras,
although there exist orthoalgebras with none; 
this assumption guarantees the existence of a tensor product of
orthoalgebras that will be defined later \cite{FoBe}.
\par
The dynamics (or physical process), which is represented as
a morphism between orthoalgebras, is discussed
 in the Heisenberg picture. 
\begin{definition}
Let $P_1$ and $P_2$ be orthoalgebras. 
A map $\phi: P_1 \to P_2$ is called a morphism if 
it satisfies the following conditions:
\begin{itemize}
\item[(i)]
For any $p,q\in P_1$ with $p \perp q$, 
$\phi(p) \perp \phi(q)$ and $\phi(p\oplus q)=\phi(p) \oplus \phi(q)$ hold. 
\item[(ii)]$\phi(1)=1$ holds. 
\end{itemize}
\end{definition}
In the Schr\"odinger picture, a state $\mu$ on $P_2$ is mapped 
onto a state $\mu\circ \phi$ on $P_1$. 
\par
We often have to consider a composite system of 
two (or more) systems. An advantage 
of treating orthoalgebra is that a tensor product 
is naturally defined in its category. 
To define the tensor product we introduce the 
notion of bimorphism. 
\begin{definition}
Let $P, Q$ and $L$ be orthoalgebras. 
A map $\beta: P\times Q \to L$ is called a bimorphism
if it satisfies the following conditions:
\begin{itemize}
\item[(i)]For $a,b \in P$ with $a\perp b$ and 
$q \in Q$, $\beta(a\oplus b, q)=\beta(a, q)\oplus \beta(b,q)$ holds. 
\item[(ii)]For $c,d\in Q$ with $c\perp d$ and $p\in P$, 
$\beta(p,c\oplus d)=\beta(p,c)\oplus \beta(p,d)$ holds.
\item[(iii)]
$\beta(1,1)=1$ holds. 
\end{itemize}
\end{definition} 
\begin{definition}
Let $P$ and $Q$ be orthoalgebras. 
$(T, \tau)$ consisting 
of an orthoalgebra 
$T$ and a bimorphism $\tau: P\times Q \to T$ 
is called a tensor product of $P$ and $Q$ if the following 
conditions are satisfied:
\begin{itemize}
\item[(i)]If $L$ is an orthoalgebra 
and $\beta: P\times Q \to L$ 
is a bimorphism, there exists a morphism $\alpha: T \to L$ 
such that $\beta =\alpha\circ \tau$.
\item[(ii)]
Every element of $T$ is a finite orthogonal sum 
of elements of the form $\tau(p,q)$ with $p\in P$ and $q\in Q$. 
\end{itemize}
We represent 
$T$ and $\tau(p,q)$ as $P\otimes Q$ and $p\otimes q$, respectively.
\end{definition}
If there exists at least one bimorphism $\beta: P\times Q \to L$ 
the above defined tensor product exists. 
The existence of the bimorphism is often confirmed by the
existence of 
a sufficiently rich set of states \cite{FoBe}, 
which is assumed 
throughout the paper.
\section{Cloning on orthoalgebras}\label{sect:main}
This section describes a cloning process on orthoalgebras.
Cloning is an operation that produces a pair of copies for
an arbitrary given state. 
Classical theory realizes this operation easily.
A simple model is illustrated as follows:
Consider a classical system having a
discrete finite sample space $\Omega_N=\{1,2,\ldots,N\}$. 
Every state can be described by a probability 
distribution $(p_n)_{n\in \Omega_N}$ on it. 
There exists an observable that distinguishes 
all the pure states. 
A composite system with
a doubled sample space $\Omega_N \times \Omega_N$ is considered. 
Measurement of the observable that perfectly distinguishes the 
pure states followed by the preparation of a pair of the identified state
clones an arbitrary state. 
That is, a state $(p_n)$ is mapped to 
a state $(p_{nm})_{(n,m)\in \Omega_N \times \Omega_N}$ 
on the composite system that is defined by 
$p_{nm}=p_n \delta_{nm}$, whose 
marginal state on 
each system coincides with the original state $(p_n)$. 
\par
The following steps describe the characterization of cloning 
operation in the algebraic setting:
Consider an orthoalgebra $P$ that 
has a separating state space. 
That is, $p=q$ follows if $p, q\in P$ satisfy $\omega(p)=\omega(q)$ for 
all the states $\omega$.
If a state $\omega$ on $P$ is cloned, 
the cloned state $\omega'$ on $P\otimes P$ must satisfy
$\omega'(q\otimes 1)=\omega'(1\otimes q)=\omega(q)$. 
We described here that states are mapped to 
other states as time evolves; in the dual picture, however,
observables are mapped backward with respect to time. 
That is, a map $\phi: P\otimes P \to P$ describes 
the time evolution. It should satisfy 
\begin{eqnarray*}
\omega\circ \phi(q\otimes 1)=\omega \circ \phi(1\otimes q)
=\omega(q)
\end{eqnarray*}
for any $q\in P$. 
If
$P$ has a separating state space, 
this condition implies that for every $q \in P$, 
\begin{eqnarray*}
\phi(q\otimes 1)=\phi(1\otimes q)=q
\end{eqnarray*}
holds.
We apply this relation as a defining property of 
a cloning map. 
\begin{definition}
Let $P$ be an orthoalgebra.
A morphism $\phi: P \otimes P\to P$ is called a cloning map
if the following conditions are satisfied:
\begin{itemize}
\item[(i)]For any $p\in P$, $\phi(p\otimes 1) =p$ holds.
\item[(ii)]For any $p \in P$, $\phi(1 \otimes p)=p$ holds.
\end{itemize}
If there exists a cloning map $\phi: P\otimes P \to P$, 
then $P$ is said to satisfy a cloning property. 
\end{definition}
The following is our main theorem: 
\begin{theorem}\label{mainortho}
Let $P$ be an orthoalgebra. 
$P$ has a cloning property if and only if 
$P$ is a Boolean algebra.
\end{theorem}
Because a Boolean algebra can be considered as 
a set of sharp propositions in a classical system 
according to Stone's representation theorem, 
this theorem essentially claims that 
if cloning operation can be performed in 
a system, the system is classical. 
To prove this theorem, we consider
a lemma with respect to the cloning map.
\begin{lemma}\label{lemma:ortho1}
Let $P$ be an orthoalgebra with a cloning property 
and $\phi:P\otimes P\to P$ be a
cloning map. The following statements are satisfied: 
\begin{itemize}
\item[(i)]
For $p,q\in P$, $\phi(p\otimes q)=0$ if and only 
if $p\perp q$.
\item[(ii)]
For any $p\in P$, $\phi(p\otimes p)=p$ holds. 
\end{itemize} 
\end{lemma}
{\bf Proof:}
Let us begin with the `if' part of (i). 
Assume that $p,q\in P$ satisfy $p\perp q$. 
Since $1=q\oplus q'$ holds, we obtain 
$p\otimes 1=p\otimes (q\oplus q')
=p\otimes q \oplus p\otimes q'$. 
Since $\phi$ is a morphism, we obtain 
\begin{eqnarray*}
p&=&\phi(p\otimes 1)
\\
&=&\phi(p\otimes q \oplus p\otimes q')
\\
&=&\phi(p\otimes q)\oplus \phi(p\otimes q').
\end{eqnarray*}
Similarly, with $1=p\oplus p'$, we obtain 
\begin{eqnarray*}
q&=&\phi(1\otimes q)
\\
&=& \phi(p\otimes q)\oplus \phi(p'\otimes q).
\end{eqnarray*}
That is, we have $p,q \geq \phi(p\otimes q)$.
In an orthoalgebra, $p\perp q$ implies
$p\wedge q=0$. 
Thus we obtain $\phi(p\otimes q)=0$.
\par
Conversely, we assume $\phi(p\otimes q)=0$ 
for some $p,q\in P$. 
We then obtain 
$p=\phi(p\otimes q')$ and $q=\phi(p'\otimes q)$.
Since $(p \otimes q' \oplus p\otimes q) \oplus 
(p'\otimes q  \oplus p' \otimes q')$ is defined
(equals $1$), 
$p\otimes q' \perp p'\otimes q$ follows. 
Therefore, due to the morphism quality of $\phi$, 
we obtain $p\perp q$. 
\par
Proof of (ii). 
Since $p\perp p'$ and $1=p\oplus p'$ hold,
$p=\phi(1\otimes p)=\phi(p\otimes p)\oplus \phi(p'\otimes p)$
follows. Applying (i),  $\phi(p'\otimes p)=0$ and 
$\phi(p\otimes p)=p$ hold.   
\hfill $\blacksquare$
\par
%
Applying this lemma, the following two lemmas are proved 
as given below: 
\begin{lemma}\label{poset}
Let $P$ be an orthoalgebra with a cloning property. 
Then $P$ satisfies coherence law. That is, 
if $x,y,z\in P$ are mutually orthogonal with each other, 
$(x\oplus y) \oplus z$ is defined.
\end{lemma}
{\bf Proof:}
As $x\oplus y$ is defined and $(x\oplus y) \oplus 
(x\oplus y)'=1$ holds, 
we have 
\begin{eqnarray*}
z=\phi(1\otimes z)=\phi((x\oplus y) \otimes z)
\oplus \phi((x\oplus y)'\otimes z).
\end{eqnarray*}
Decomposing the first term in the right-hand side, 
we obtain 
\begin{eqnarray*}
z=\phi(x\otimes z)\oplus \phi(y\otimes z)
\oplus \phi((x\oplus y)'\otimes z)
=\phi((x\oplus y)' \otimes z),
\end{eqnarray*}
where Lemma \ref{lemma:ortho1} (i) is used.
On the other hand, $x\oplus y =\phi((x\oplus y) \otimes 1)$
holds and $((x\oplus y) \otimes 1) \oplus ((x\oplus y)' \otimes z)$
is defined. Thus, due to the 
morphism quality of $\phi$, $(x\oplus y)\oplus z$ is defined. 
\hfill $\blacksquare$
\begin{lemma}
Let $P$ be an orthoalgebra with a cloning property. 
Then any two elements are compatible. 
That is, 
for any $p,q\in P$, there exist 
mutually orthogonal elements $a,b,r \in P$
such that $p=r\oplus a$ and $q=r\oplus b$ hold. 
In addition, this decomposition
is
unique.
\end{lemma}
{\bf Proof:}
Substituting $r:=\phi(p\otimes q)$, we obtain
\begin{eqnarray*}
p&=&\phi(p\otimes 1)=r\oplus \phi(p\otimes q'),
\\
q&=&\phi(1\otimes q)=r\oplus \phi(p'\otimes q).
\end{eqnarray*}
Substituting $a=\phi(p\otimes q')$ and $b=\phi(p'\otimes q)$, 
we obtain $p=r\oplus a$ and $q=r\oplus b$.
Since $p\otimes q'\perp p' \otimes q$ holds, 
$a\perp b$ follows. 
%
\par
If 
$p=r\oplus a$ and $q=r\oplus b$
are decompositions 
with mutually orthogonal $r,a,b$, 
we obtain 
\begin{eqnarray*}
\phi(p\otimes q)=\phi((r\oplus a) \otimes (r\oplus b))
=\phi(r\otimes r)=r,
\end{eqnarray*}
where Lemma \ref{lemma:ortho1} is used. 
Thus the decomposition is unique.
\hfill $\blacksquare$
\par
{\bf Proof of Theorem \ref{mainortho}}
\par
The above two lemmas prove that 
every orthoalgebra with the cloning property satisfies 
the coherence law and the compatibility condition and is a
Boolean algebra.
\par
Conversely, let $P$ be a Boolean algebra. 
We define $\phi: P\otimes P\to P$ by 
$\phi(p\otimes q)=p\wedge q$ and its natural extension
to their orthogonal sums. 
Its well-definedness can be proved as follows. 
Let $\{p_m,q_m,r_n,s_n\}\subset P$ be a finite family 
satisfying 
$\bigoplus_m p_m \otimes q_m =\bigoplus r_n \otimes s_n$. 
Since $P$ is Boolean, there exist finite mutually orthogonal 
nonvanishing elements $\{x_{i}\}_{i=1}^N \subset P$ and 
subsets $S_{p_m}, S_{q_m}, S_{r_n}, S_{s_n} \subset 
\{1,2,\ldots,N\}$, such that the following are satisfied:
\begin{eqnarray*}
p_m &=&\bigoplus_{i \in S_{p_m}} x_i
\\
q_m &=&\bigoplus_{i \in S_{q_m}} x_i 
\\
r_n &=& \bigoplus_{i \in S_{r_n}} x_i 
\\
s_n &=& \bigoplus_{i \in S_{s_n}} x_i .
\end{eqnarray*}
The condition $\bigoplus_m p_m \otimes q_m =\bigoplus r_n \otimes s_n$
indicates the followings:
\begin{itemize}
\item[(i)] $(S_{p_m} \times S_{q_m}) \cap (S_{p_{m'}} \times S_{q_{m'}})
=\emptyset$ for $m\neq m'$ and  
$(S_{r_n} \times S_{s_n}) \cap (S_{r_{n'}} \times S_{s_{n'}})
=\emptyset$ for $n\neq n'$ hold. 
\item[(ii)]
$\bigcup_m (S_{p_m} \times S_{q_m}) =\bigcup_n (S_{r_n} \times S_{s_n})$
holds. We denote $F:=\bigcup_m (S_{p_m} \times S_{q_m})$. 
\end{itemize}
Applying 
$p_m \wedge q_m =\oplus_{(i,i) \in S_{p_m}\times S_{q_m}}x_i$,
%
$\bigoplus_m p_m \wedge q_m
=\oplus_{(i,i)\in F} x_i$ holds, which equals 
$\bigoplus_n r_n \wedge s_n$. 
Thus $\phi$ is well-defined. 
It can be noted that this $\phi$ satisfies the conditions of 
the cloning map.
\hfill $\blacksquare$
\par
If an orthoalgebra $P$ admits an orthoalgebra $L$ and 
a bimorphism $\beta:P\times P \to L$ satisfying the following conditions, 
then $P$ is a Boolean algebra. 
\begin{itemize}
\item[(i)] There exists a morphism $\phi: L\to P$.
\item[(ii)] 
For every $p\in P$, $\phi\circ \beta(p,1)=\phi\circ \beta(1,p)=p$ 
holds.
\end{itemize}
In fact, according to the definition of $P\otimes P$, 
there exists a morphism $\alpha :P\otimes P \to L$ 
satisfying $\beta(p,1)=\alpha(p\otimes 1)=p$ and 
$\beta(1,p)=\alpha(1\otimes p)=p$ for every $p$, and 
Theorem \ref{mainortho} can be applied.
It may be worth noting that
some important examples such as $P=P({\cal H})$ and 
$L=P({\cal H}\otimes {\cal H})$ for a Hilbert space ${\cal H}$
can be treated in this manner.
\section{Cloning on effect algebras}\label{sect:effect}
In the previous section, we 
proved that if an orthoalgebra 
has a cloning property, then it is a Boolean 
algebra. 
This section considers a possible extension of 
this result to 
effect algebras.
Let us begin with the definition of an effect algebra. 
\begin{definition}
Let us consider $(L,0,1,\oplus)$ consisting of a set 
$L$ which contains two special elements $0$ and $1$ and 
a partially 
defined binary operation $\oplus$.
If 
the quadruple 
satisfies the following conditions for 
all $p,q,r\in L$, 
then $(L,0,1,\oplus)$ is called an effect algebra: 
\begin{itemize}
\item[(i)]
If $p\oplus q$ is defined (denoted by $p \perp q$), then 
$q\oplus p$ is also defined and $p\oplus q =q\oplus p$. 
\item[(ii)]
If $q\perp r$ and $p\perp (q\oplus r)$ hold, 
then $p\perp q$ and $(p\oplus q) \perp r$, and
$p\oplus (q\oplus r)=(p\oplus q) \oplus r$ hold. 
\item[(iii)]
 For every $p\in L$, there exists a unique $q\in L$ 
 such that $p\perp q$ and $p\oplus q =1$ hold. 
 We represent such unique $q$ as $p'$. 
\item[(iv')]
If $p\perp 1$, then $p=0$.
\end{itemize}
\end{definition}
It may be noted that the condition (iv) in Definition
\ref{def:ortho} is stronger than 
(iv'). That is, every orthoalgebra is an effect algebra; 
the converse is not true. 
A partial order $\leq$ is introduced as in the orthoalgebra. 
While every element of an orthoalgebra is sharp, that is, 
$p\wedge p'=0$ holds,  
elements of an effect algebra may not be sharp.
If all the elements of an effect algebra 
are sharp, then this algebra turns out to be an orthoalgebra. 
\begin{example}
Let us consider a quantum system described by a Hilbert space
${\cal H}$. $E({\cal H})$ is defined as a set 
of all the positive operators $x$ on ${\cal H}$ that satisfy 
$x\leq {\bf 1}$, where ${\bf 1}$ is an identity operator.  
$x \oplus y=x+y$ (summation as operators) 
is defined if $x+y\leq {\bf 1}$ holds. 
$0$ and $1$ are the null operator and identity operator, respectively. 
The quadruple $(E({\cal H}),0,1,\oplus)$ 
becomes an effect algebra. 
\end{example}
The notions of state, 
dynamics and tensor product 
are defined by simply replacing orthoalgebra 
by effect algebra in the previous definitions. 
Cloning condition is defined as in the orthoalgebra. 
\begin{definition}
Let $L$ be an effect algebra. 
$\phi: L\otimes L \to L$ is called a cloning map if 
and only if it satisfies $\phi(p\otimes 1)=\phi(1\otimes p)=p$
for any $p \in L$. 
An effect algebra for which there exists a cloning map is 
said to satisfy a cloning property.
\end{definition}
The property stated in Lemma \ref{lemma:ortho1}, 
in contrast to that in orthoalgebras wherein it played 
a crucial role, 
does not hold in effect algebras. 
Certain partial results on effect algebras are 
elucidated as follows. 
The first result is related to atomic effect 
algebras. 
\begin{definition}
Let $L$ be an effect algebra. 
A non-zero element $a\in L$ is called an atom if and only if 
$[0,a]:=\{x|0\leq x\leq a\}=\{0,a\}$ holds.
\end{definition}
\begin{definition}
Let $L$ be an effect algebra. 
If for every non-zero $p\in L$ there exists an atom $a\in L$ satisfying 
$a\leq p$, then $L$ is called an atomic algebra. 
\end{definition}
That is, an atomic algebra has `smallest' units in it. 
On the other hand, the `boundedness' of the algebra 
is imposed by another condition. 
\begin{definition}
Let $L$ be an effect algebra. 
For every element $x\in L$, we define its isotropic index $\iota(x)$
as the maximal nonnegative integer $n$ such that $na:=a\oplus a\oplus
\cdots \oplus a$ ($n$ times) is defined. 
If $\iota(x)$ is finite for every $x\neq 0$, then 
$L$ is called an Archimedean algebra. 
\end{definition} 
The following is the first result on effect algebras.
\begin{theorem}
If an atomic Archimedean effect algebra $L$ has a cloning 
property, $L$ is a Boolean algebra. 
\end{theorem}
{\bf Proof:}
First let us prove that every atomic element is a sharp element. 
Suppose $a$ is an atom. 
As $[0,a]=\{0,a\}$, if $a\leq a'$ holds then $a\wedge a'=a$, 
otherwise $a\wedge a'=0$. 
Assume that $a\leq a'$ holds ($a\perp a$ follows), 
then $\phi(a\otimes a')\geq \phi(a\otimes a)$ holds. 
Since $a=\phi(a\otimes 1)
\geq \phi(a\otimes a)$ holds, 
$\phi(a\otimes a)$ is $a$ or $0$. 
If $\phi(a\otimes a)=a$ holds, then it implies 
$ \phi(a\otimes a')\geq a$. 
Then the cloning property leads to
\begin{eqnarray*}
a=\phi(a\otimes 1)=\phi(a\otimes a')\oplus 
\phi(a\otimes a)\geq 2a. 
\end{eqnarray*}
Therefore implying $a=0$ which leads to a contradiction. 
Thus $\phi(a\otimes a)=0$ should hold
and $\phi(a\otimes a')=a$ follows. 
Since we assumed $a\perp a$, 
$2a=a\oplus a$ 
is defined and $\phi(2a \otimes a')=2\phi(a\otimes a')=2a$ holds. 
As $a'\geq \phi(2a\otimes a')$ holds, it means
that $a'\geq 2a$ should hold;
thus $3a \in L$ is defined. 
Repeating the same arguments, 
we obtain $Na\in L$ for arbitrarily large $N$. 
As $L$ is Archimedean, $a=0$ follows. This 
contradicts the nonvanishing characteristics of $a$. 
Thus it can be concluded that $a\wedge a'=0$ for every 
atom $a$. 
\par
Consider an arbitrary element $p\in L$. 
Assume a non-zero element $x\in L$ 
satisfying $x\leq p,p'$. 
As $L$ is atomic, there exists an atom $a\leq x$; 
thus $a\leq p,p'$ holds, 
implying $a'\geq p,p'$
and 
$a\leq p\leq a'$.
It indicates $a\wedge a'=a$. 
We, however, proved that it does not hold
for an atom $a$; therefore, this leads to a contradiction.
Hence it can be concluded 
that $x=0$ 
and $p\wedge p'=0$. 
Thus all the elements in $L$ are sharp, 
which means that $L$ is an orthoalgebra. 
According to Theorem \ref{mainortho}, 
an orthoalgebra with cloning property is a
Boolean algebra. 
\hfill $\blacksquare$
\par
In non-atomic effect algebras, 
the above theorem does not hold. 
The following is a simple example. 
$\Omega_N:=\{1,2,\ldots,N\}$ 
is a finite discrete set consisting of $N$ points.
Let $L_N:=[0,1]^{\Omega_N}=
\{f|\Omega_N \to [0,1]\}$ be the set of 
all functions from $\Omega_N$ to $[0,1]$. 
On $L_N$, a partial binary operation $\oplus$ 
can be defined by $(f\oplus g)(x)=f(x)+g(x)$ for all $x\in \Omega_N$ if 
$f(y)+g(y)\in [0,1]$ for all $y \in \Omega_N$ holds. 
Both $0$ and $1$ are defined in a natural manner. 
It can be noted that $L_N$ becomes 
an effect algebra although it is not an orthoalgebra. 
In addition, it can be proved that 
$L_N \otimes L_N$ is isomorphic to 
$L_{N^2}:=\{f|\Omega_N \times \Omega_N \to [0,1]\}$. 
A cloning map can be defined by 
\begin{eqnarray*}
\phi(f)(x):=f(x,x)
\end{eqnarray*}
for $f\in L_{N^2}$. 
It may be worth noting that the sharp elements of this 
effect algebra forms a Boolean algebra. 
\par
Based on the above example, it may be expected that 
the cloning property on effect algebras is related to
 the classicality.
We now explain the definition of hidden variable 
introduced by Pulmannov\'a \cite{Pulmannova}. 
It uses the following MV-algebra \cite{Chang}: 
\begin{definition}
Let $M$ be a set with two special elements $0$ and $1$. 
If on $M$ a binary operation $+$ and 
a unary operation $'$ are defined and satisfy the following conditions
for all $a,b,c\in M$, then 
$(M,+,\ ', 0,1)$ is called an MV-algebra. 
\begin{itemize}
\item[(i)]
$a+b=b+a$
\item[(ii)]
$(a+b)+c=a+(b+c)$
\item[(iii)]
$a+a'=1$
\item[(iv)]
$a+0=a$
\item[(v)]
$(a')'=a$
\item[(vi)]
$0'=1$
\item[(vii)]
$a+1=1$
\item[(viii)]
$(a'+b)'+b=(a+b')'+a$.
\end{itemize}
\end{definition} 
If we define a partial binary operation $\oplus$ by 
$a\oplus b :=a +b$ only for $a, b\in M$ with $a\leq b'$, 
then $(M,0,1,\oplus)$ becomes an effect algebra. 
The states on an MV-algebra are defined by the 
states on its corresponding 
effect algebra. 
\begin{definition}\label{def:hidden}(\cite{Pulmannova})
Let $L$ be an effect algebra. We consider that $L$ admits hidden 
variables if and only if there exists an MV-algebra, $M$ satisfying
the following conditions.
\begin{itemize}
\item[(i)] There exists a morphism 
from $L$ to $M$, that is, a map 
$h: L\to M$ such that 
for all $p,q\in L$ with $p\perp q$, $h(p\oplus q)=
h(p)+ h(q)$ holds. 
\item[(ii)] For every state $\omega$ on $L$, there exists a 
state $\overline{\omega}$ on $M$ such that 
$\overline{\omega}\circ h(q)=\omega(q)$ for every $q\in L$ holds. 
\end{itemize}
\end{definition}
The following theorem is a partial result on 
non-atomic effect algebras. 
\begin{theorem}\label{saigo}
Let $L$ be an 
effect algebra with a cloning property. 
Suppose there exists a family $\{p_1,p_2,\ldots,p_N\} 
\subset L$ such that 
$[0,p_n]$ is a linearly ordered ideal for every $p_n$ and 
$1=\oplus_{n=1}^N p_n$ holds. 
$L$ admits a hidden variable. 
\end{theorem}
{\bf Proof:}
We define a binary operation $+$ on
$[0,p_n]$ by $x+y=x\oplus y$ for $x,y \in [0,p_n]$ with $x\perp y$ 
and $x+y =p_n$ otherwise.
A unary operation $'$ is defined by $x':=p_n \ominus x$ for 
$x\in [0,p_n]$. (That is, $x'$ is a unique element that satisfies
$x\oplus x'=p_n$.) 
Then it can be noted that $([0,p_n],+,',0,p_n)$ becomes an 
MV-algebra. 
Let us consider their Cartesian product, $\Pi_n [0,p_n]$, 
which becomes again an MV-algebra by defining the summation 
and the unary operation 
`pointwise' and $1:=(p_n)_{n=1,\ldots,N}$. That is, 
the summation is defined as $(x_n)_{n=1,\ldots,N} +(y_n)_{n=1,\ldots,N} 
:=(x_n +y_n)_{n=1,\ldots,N}$, and the unary operation 
is defined as $(x_n)'_{n=1,\ldots,N}
:=(x_n')_{n=1,\ldots,N}$. We define a map $h: L\to M$
by $h(x)=(\phi(p_n \otimes x))_{n=1,\ldots,N}$.
$h(x)=h(y)$ means $\phi(p_n\otimes x)=\phi(p_n \otimes y)$ for 
all $n$. As $x=\phi(1\otimes x)=\oplus_{n=1}^N \phi(p_n \otimes x)$ 
holds, it implies $x=y$. 
In addition, for any $(x_n)_{n=1,\ldots,N}$ 
satisfying $x_n \in [0,p_n]$ for each $n$, 
$x:=\oplus_{n=1}^N x_n$ is defined and it satisfies 
$\phi(p_n \otimes x)=\phi(p_n \otimes x_n)=x_n$ for each $n$ 
since $p_n$ is a 
sharp element (Lemma 1.9.6 in \cite{TheBook}) and $p_n \wedge x_m=0$
holds for $n\neq m$. 
Thus $h$ is a bijection. 
\par
If a pair $x,y \in L$ satisfies $x \perp y$, 
$x\oplus y=\oplus_{n=1}^N \phi(p_n \otimes (x\oplus y))$
holds. Therefore $h(x\oplus y)= (\phi(p_n \otimes (x \oplus y))_{n=1,\ldots,N}
=(\phi(p_n \otimes x)+\phi(p_n \otimes x))_{n=1,\ldots,N}
=h(x)+h(y)$ follows. 
That is, $h$ satisfies the condition (i). 
Conversely, if $h(x) \leq h(y)'$ holds, 
it means $\phi(p_n \otimes x) \oplus \phi(p_n \otimes y)
\leq p_n$ for each $n$. 
It entails $\left( \oplus_{n=1}^N \phi(p_n \otimes x)\right) 
\oplus \left( \oplus_{n=1}^N
\phi(p_n \otimes y)
\right)\leq 1$. 
That is, $x \leq y'$ holds. 
Thus we proved that $x \leq y'$ if and only if 
$h(x) \leq h(y)'$. 
Let $\omega$ be a state on $L$. 
A state on $M$, 
$\overline{\omega}$, is defined by 
$\overline{\omega}(\{x_n\})=\omega(\oplus_{n=1}^N x_n)$.
This $\overline{\omega}$ satisfies 
the condition (ii).
%
\hfill $\blacksquare$
\section{Summary}
This paper considers the no-cloning theorem on
orthoalgebras and effect algebras. We proved that 
an orthoalgebra admits cloning operation if and only if it is 
a Boolean algebra. That is, 
cloning operation can be performed only on classical systems. 
In addition, we 
proved that an atomic Archimedean effect algebra 
with a cloning property is a Boolean algebra. 
We also 
obtained a partial result
that indicates a connection between 
the cloning property and hidden variables. 
Although we conjecture that effect algebra with 
the cloning property admits 
a hidden variable, 
we have not succeeded in proving it. 
\par
{\bf Acknowledgements}
\par
The authors thank an anonymous referee for fruitful comments.

\end{document}